# Direct observation of site-specific dopant substitution in Si doped $(Al_xGa_{1-x})_2O_3$ via Atom Probe Tomography

Jith Sarker[1], A F M Anhar Uddin Bhuiyan[2], Zixuan Feng[2], Hongping Zhao[2,3], and Baishakhi Mazumder[1]*

[1]Department of Materials Design and Innovation, University at Buffalo, Buffalo, NY 14260, USA
[2]Depertment of Electrical and Computer Engineering, The Ohio State University, Columbus, OH 43210, USA
[3]Department of Materials Science and Engineering, The Ohio State University, Columbus, OH 43210, USA

E-mail: baishakh@buffalo.edu

## Abstract

In this work, the interaction of n-type dopants in Si doped $(Al_xGa_{1-x})_2O_3$ films with varying Al content over the entire composition range (x = 0-100%) was analyzed using atom probe tomography. An almost uniform dopant distribution with dopant density in the range of $10^{18}$ cm$^{-3}$ was obtained in all $(Al_xGa_{1-x})_2O_3$ layers containing different Al contents. We have demonstrated that for the single phase $\beta$-$(Al_xGa_{1-x})_2O_3$ films with Al content of x<0.30, dopants prefer to occupy on Ga sites while Al site is preferred for high Al content (x>0.50) $(Al_xGa_{1-x})_2O_3$ layers. It was also observed for Al content, x = 0.30-0.50, no specific cationic site occupancy was observed, Si occupies either Al or Ga sites. This can be attributed to highly inhomogeneous layers within this composition range due to which dopant Si atoms are either in the Al-rich or Al-depleted regions.

## 1. Introduction

$(Al_xGa_{1-x})_2O_3$ is a thermodynamically stable ultra-wide bandgap semiconductor with a high predicted breakdown field and higher Baliga's figure of merit (BFoM) making it a rising candidate for future high power switching devices and deep ultra-violet (DUV) optical applications [1-3]. $(Al_xGa_{1-x})_2O_3$ has already demonstrated its potential in high speed field effect transistors (FETs) [4], Schottky barrier diodes (SBDs) [5] and DUV photodetectors [6]. For these devices to perform up to their theoretical limit, extrinsic n-type doping is necessary since intrinsic defects i,e, oxygen vacancies do not significantly contribute to conduction [7,8]. Si, Sn, and Ge are being considered as suitable dopants for $(Al_xGa_{1-x})_2O_3$ inspired by the high dopant density as well as controlled doping profile achieved in the case of $Ga_2O_3$ based devices [8-10]. Although, Si doping in $(Al_xGa_{1-x})_2O_3$ has been reported by Bhuiyan et. al. [11], Ranga et. al. [12] and Hassa et. al. [8] very recently, doping of $(Al_xGa_{1-x})_2O_3$ films is still in its infancy and needs adequate understanding.

Impurity doping is often associated with the formation of specific native defects and these defects are more likely to diffuse towards the active bulk region [13]. In $Ga_2O_3$, it was observed that Si substitutes on the Ga site and contributes to shallow or deep level defect states [7] as well as structurally complex defects [14]. Similar Si occupancy on the Ga site ($Si_{Ga}$) was reported for GaAs and GaN with the formation of a complex defect with $V_{Ga}$, ($V_{Ga}$-$Si_{Ga}$) [15,16]. In AlGaN, although Si can occupy either Ga ($V_{Ga}$) or Al ($V_{Al}$) site [16], at higher Al content of >60%, $V_{III}$-$Si_{III}$ defect complexes on Al site ($V_{Al}$-$Si_{III}$) are more common owing to the increasing amount of $V_{Al}$ formation [13,17]. These defects induced by Si doping are electrically active [18] and can directly affect the electrical and optical performance of the fabricated devices via acting as a charge trapping center. Electron concentration, hence carrier mobility increases proportionally with Si concentration at low doping level (up to $10^{17}$ cm$^{-3}$), while at higher doping level ($\geq 10^{18}$ cm$^{-3}$), a decrease in free electron concentration (carrier mobility) is observed due to a variety of compensating defects formation resulting from cationic (Ga or Al site) substitution [19]. This phenomenon is termed as the compensation knee. The mechanism and specification of the predominant defects ($V_{Ga}$-$Si_{Ga}$ or $V_{Al}$-$Si_{Al}$) responsible for this compensation knee, *i.e.* reduced carrier mobility with high doping level, have not been conclusively understood yet for $(Al_xGa_{1-x})_2O_3$. Therefore, it is critical to have a clear understanding of how the n-type Si doping interacts with $(Al_xGa_{1-x})_2O_3$ matrix in order to realize high mobility devices.

To gain a deep insight into dopant interaction in $(Al_xGa_{1-x})_2O_3$ films, atomic level investigation of doping chemistry is desired. Atom probe tomography (APT) is an advanced nanoscale characterization tool enabling 3D visualization of constituent elements atom by atom with excellent spatial and chemical resolution [20]. APT is capable of providing information of each dopant's locations and atomic distribution in the neighbouring sites in the host materials [21] that can be used to frame specific site occupancy by individual dopant atoms and would be crucial to account for the compensation effect resulting from cation specific defect complex like $V_{III}$-$Si_{III}$. In this paper, we employed APT to study dopant's interaction in Si doped $(Al_xGa_{1-x})_2O_3$ films with varying Al content over the whole composition range of x = 0-1.0. A statistical analysis method, radial distribution function, was adapted to elucidate the mechanism of Si dopant's incorporation in $(Al_xGa_{1-x})_2O_3$ films. Since high Al content in $(Al_xGa_{1-x})_2O_3$ is required for bandgap enlargement, a critical understanding of subsequent dopant's interaction in these films at different Al contents will be significant for realization of high performance $(Al_xGa_{1-x})_2O_3$ based devices.

## 2. Experimental Section

The Si doped $(Al_xGa_{1-x})_2O_3$ heterostructure with varying Al content, x = 0, 0.10, 0.20, 0.30, 0.40, 0.50, 0.60, 0.80 and 1.0 with a thickness of 20 nm each were grown by metal organic chemical vapor deposition (MOCVD) on Fe-doped semi-insulating (010) β-$Ga_2O_3$ substrates (from Novel Crystal Technology, Inc.). Details of the growth can be found elsewhere [11,22]. For APT specimen preparation, an additional $(Al_{0.30}Ga_{0.70})_2O_3$ layer (50 nm) was deposited as a sacrificial layer on top of the $(Al_xGa_{1-x})_2O_3/Ga_2O_3$ heterostructures. The Focused ion beam (FIB) milling method was used to prepare needle-shaped specimens following standard lift-out and annular milling procedure as reported by Thompson et. al. [23]. APT experiment was performed using pulsed laser assisted CAMECA Local Electrode Atom Probe (LEAP 5000X HR) system at a specimen base temperature of 50 K under ultrahigh vacuum of $<10^{-11}$ mbar. The controlled field evaporation of atoms from the needle-shaped specimen was achieved by employing a UV laser at a pulse energy of 20 pJ. The pulse frequency was 200 kHz and the detection rate was 0.005 atoms per pulse. CAMECA Integrated Visualization and Analysis Software (IVAS, version 3.6.14) was used for 3D reconstruction of the specimen and advanced data analysis. Statistical analysis was conducted employing radial distribution function (RDF) over small bulk volumes ($10\times10\times4$ nm$^3$) in each $(Al_xGa_{1-x})_2O_3$ layers to investigate the nearest neighbor distribution of elements in order to understand the dopant's incorporation mechanism in these films.

## 3. Results

Figure 1 (a) illustrates the schematic diagram of the $(Al_xGa_{1-x})_2O_3$ heterostructures with different Al content of x = 0-100% layers along the growth direction. To visualize the dopant incorporation within the heterostructures, lateral Si distribution on XY plane, perpendicular to the growth direction was plotted for 4 nm thick volumes extracted from the bulk regions of each individual layer as depicted in Figure 1 (a-j). The actual sizes of the extracted volumes from each layer are provided in table I. From the initial visualization of lateral Si distribution in each $(Al_xGa_{1-x})_2O_3$ layer in Figure 1 (b-j), it is confirmed that a closely random dopant distribution is obtained with a dopant concentration of $\sim$1-7$\times 10^{18}$ cm$^{-3}$. The measured dopant concentration closely resembles with the concentration range expected from the average dopant incorporation of $\sim$1-5$\times 10^{18}$ cm$^{-3}$ during the epi-growth. This implies, desired dopant incorporation with an excellent dopant profile is achieved. However, in the lateral Si distribution plots, several small island-like high Si concentration regions are observed. These artifacts may arise from local density variation of the doped $Al_xGa_{1-x})_2O_3$ matrix, since Al/Ga ratio is varying, any fluctuation of local stoichiometry [22] could result in an apparent non-uniform Si distribution. Also, the Si count being very low in this case, the vertical projection of two or more Si atoms in the same or closely located XY planes (two Si ions overlaid on each other) would contribute in high density regions in lateral Si distribution as evident by Fig. 1(i) and 1(j). Therefore, additional statistical analysis is necessary to determine the uniformity of dopant distribution.

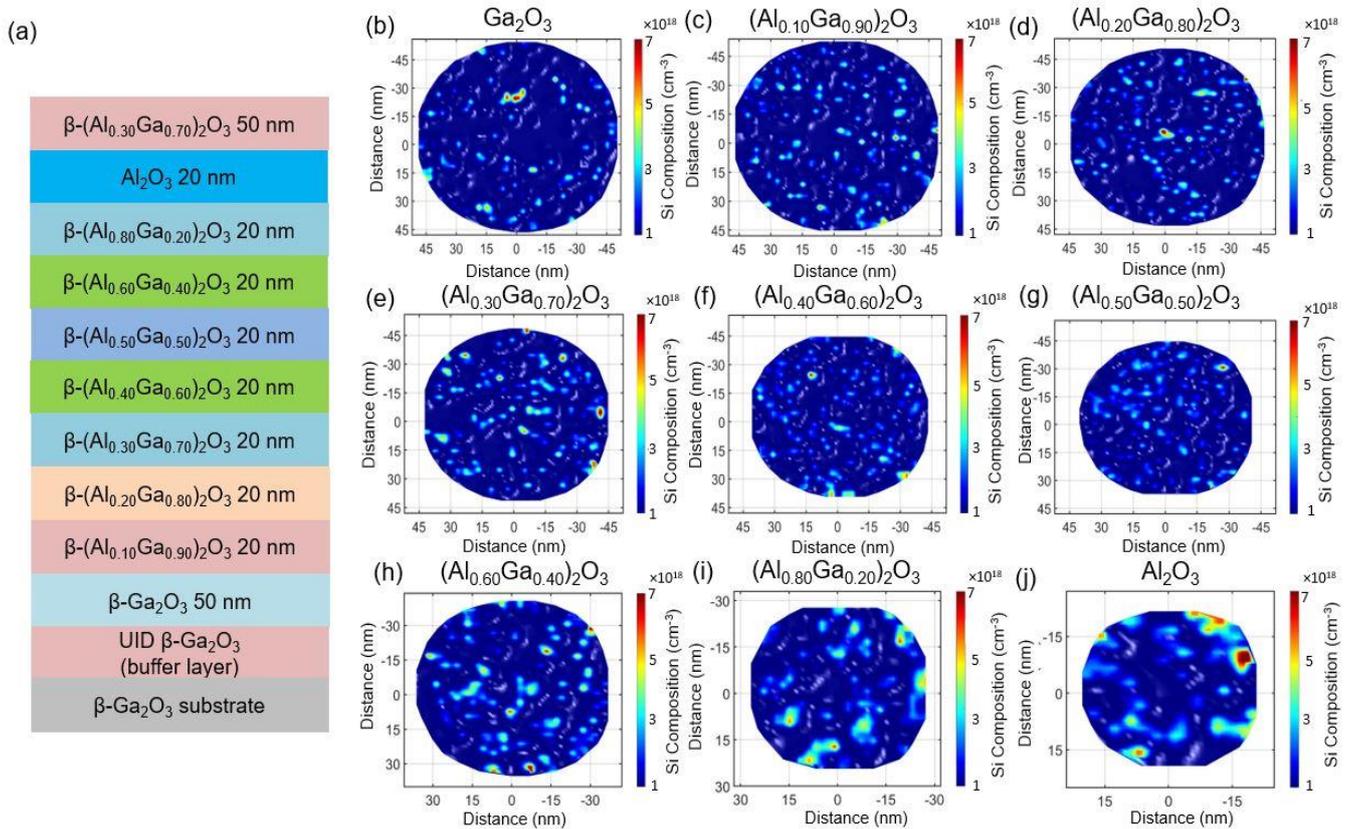

Figure 1 (a). A schematic diagram of the Si doped layered $(Al_xGa_{1-x})_2O_3$ heterostructure with varying Al composition. Lateral distribution Si distribution in each layer with Al composition of (b) x = 0.0, (c) x = 0.10, (d) x = 0.20, (e) x = 0.30, (f) x = 0.40, (g) x = 0.50, (h) x = 0.60, (i) x = 0.80 and (j) x = 1.0, showing that Si incorporation is as expected within the range of $10^{18}$ cm$^{-3}$ and dopants are randomly distributed.

Table I. Region of interest (ROI) volumes extracted from the bulk region of each layer with different Al composition to observe lateral Si distribution

| Layers | Volume of the extracted region (nm$^3$) |
|---|---|
| $Ga_2O_3$ | $96\times96\times4$ nm$^3$ |
| $(Al_{0.10}Ga_{0.90})_2O_3$ | $94\times94\times4$ nm$^3$ |
| $(Al_{0.20}Ga_{0.80})_2O_3$ | $92\times92\times4$ nm$^3$ |
| $(Al_{0.30}Ga_{0.70})_2O_3$ | $90\times90\times4$ nm$^3$ |
| $(Al_{0.40}Ga_{0.60})_2O_3$ | $85\times85\times4$ nm$^3$ |
| $(Al_{0.50}Ga_{0.50})_2O_3$ | $80\times80\times4$ nm$^3$ |
| $(Al_{0.60}Ga_{0.40})_2O_3$ | $70\times70\times4$ nm$^3$ |
| $(Al_{0.80}Ga_{0.40})_2O_3$ | $55\times55\times4$ nm$^3$ |
| $Al_2O_3$ | $45\times45\times4$ nm$^3$ |

Figure 2(a) depicts the APT reconstructed 3D volume of the graded $(Al_xGa_{1-x})_2O_3$ heterostructure highlighting Al distribution along with the layers with x = 0-100%. In this representation, a temperature color map was used to observe the Al concentration variation along the heterostructure with red representing the highest Al concentration (x = 1.0) while lowest Al concentration (x = 0) is shown in blue. This color temperature variation enables to identify each layer with different Al contents. The detailed analysis for alloy composition variation for this similar structure is reported in our earlier work [22]. Each $(Al_xGa_{1-x})_2O_3$ layer was investigated individually to verify the Si distribution whether the dopants are forming any cluster or not. A statistical method, frequency distribution analysis (FDA) of Si distribution was conducted from the same bulk volumes used for lateral Si distribution in Figure 1 (b-j) using a bin size of 300 atoms. The total number of bins in each volume taken in each $(Al_xGa_{1-x})_2O_3$ layers is provided in table II. In FDA, the deviation of observed elemental distribution from a binomial fitting represents the presence of elemental segregation or inhomogeneity in the layer [24]. The FDA results for Si in $(Al_xGa_{1-x})_2O_3$ films with x = 0-100% are shown in Figure 2 (b-j). It is observed that, for each case, the observed Si distribution closely resembles the binomial fit for random (uniform) elemental distribution. The Pearson coefficient, μ in each case is low (<0.1). Pearson co-efficient (μ) tending to 1 implies statistically significant elemental segregation [25]. The very low μ-values suggest no statistically considerable Si segregation is present. The null hypothesis testing of Si distribution provides low P-value (0.05-0.1) which suggests, the null hypothesis can be rejected at a 90-95% confidence level and Si has a distribution which is statistically similar to a binomial random distribution.

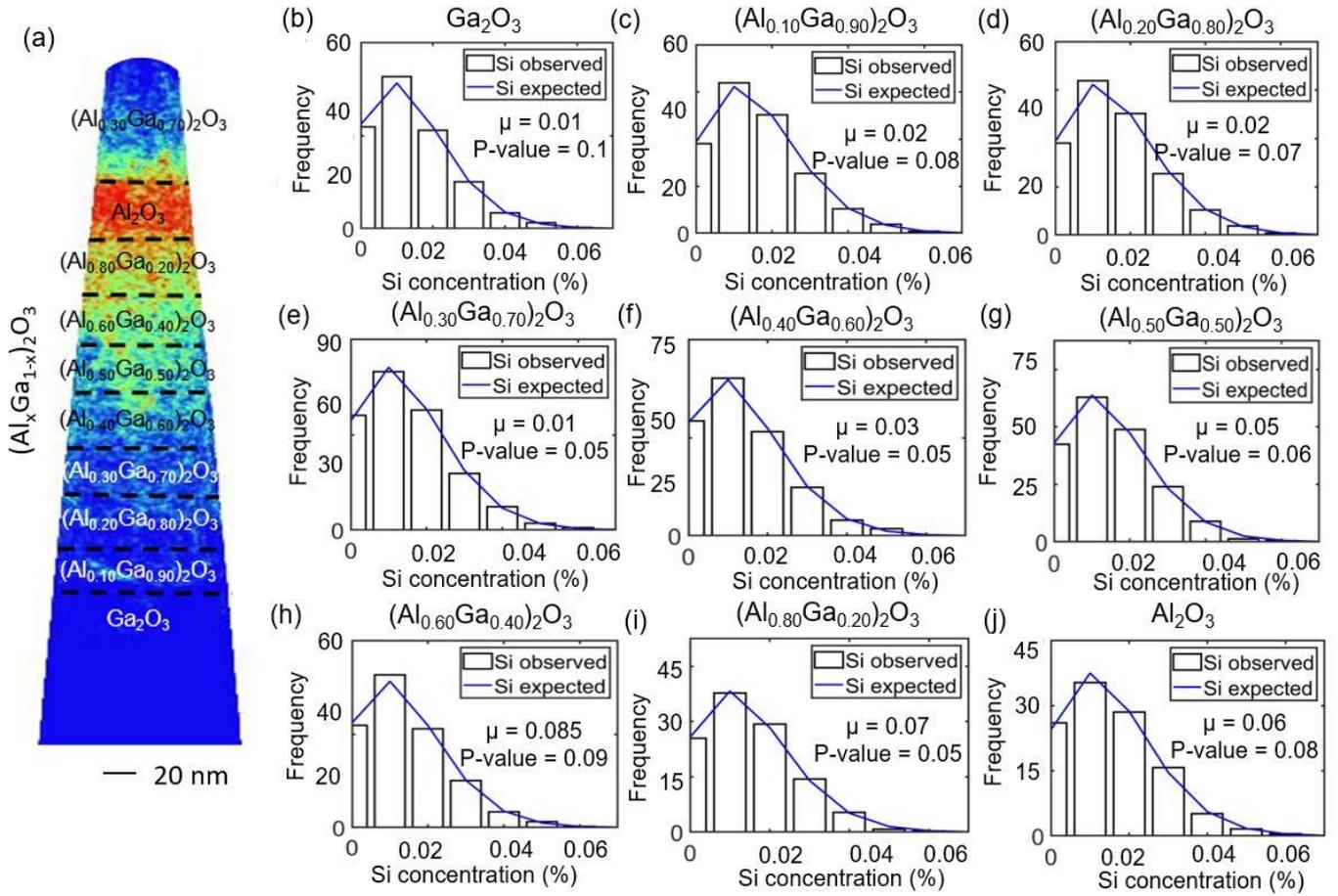

Figure 2 (a) Volume render of $(Al_xGa_{1-x})_2O_3$ heterostructure with x = 0-100% highlighting Al distribution within the APT data set, (b-i). Frequency distribution analysis for Si distribution in $(Al_xGa_{1-x})_2O_3$ layers with x = 0-100%, showing statistically uniform Si distribution in all these layers.

Table II. Total number of bins in FDA analysis for each layer using a bin size of 300 atoms

| Layers | Total number of bins |
|---|---|
| $Ga_2O_3$ | 5225 |
| $(Al_{0.10}Ga_{0.90})_2O_3$ | 5120 |
| $(Al_{0.20}Ga_{0.80})_2O_3$ | 4833 |
| $(Al_{0.30}Ga_{0.70})_2O_3$ | 4666 |
| $(Al_{0.40}Ga_{0.60})_2O_3$ | 4333 |
| $(Al_{0.50}Ga_{0.50})_2O_3$ | 3800 |
| $(Al_{0.60}Ga_{0.40})_2O_3$ | 2166 |
| $(Al_{0.80}Ga_{0.40})_2O_3$ | 1510 |
| $Al_2O_3$ | 985 |

As the alloy composition is changing substantially, it is crucial to understand the mechanism at which dopants are interacting with the matrix. A statistical analysis method, RDF was applied to the APT data to understand the dopant chemistry within the alloy matrix. RDF determines the radial concentration profile starting from each atom detected for any specifically chosen species and returns the probability density of finding an atom $j$ at a distance $r$ given that an atom $i$ being the center [26]. RDF normalizes the local concentration of a selected species to the bulk concentration within the volume

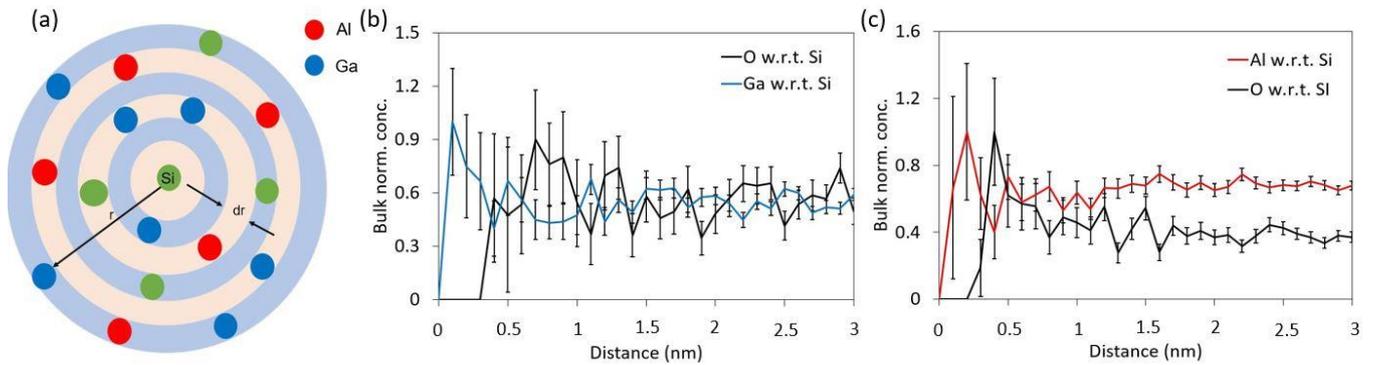

Figure 3 (a). A schematic diagram illustrating the operation of radial distribution function; RDF result on dopant Si atom in (b) $Ga_2O_3$ showing Si is occupying in Ga site, (c) $Al_2O_3$ showing Si is occupying in Al site. In both case, O concentration surrounding the center Si atoms position is zero indicating cationic substitution by the dopants.

selected. The RDF function works in a radially outward direction from each center atom of interest and the measured bulk concentration is averaged across the sample [27]. Figure 3(a) illustrates the operation of RDF in this study where each Si is considered as a center atom and starting from that origin, the bulk normalized concentration of other detected species (Al, Ga, and O) were measured. Figure 3(b) and 3(c) illustrates the RDF analysis of $Ga_2O_3$ (x = 0%) and $Al_2O_3$ (x = 100%), showing bulk normalized concentration (BNC) for Ga and Al, respectively with BNC of O as reference considering Si atoms at the origin (distance, d = 0). The BNC for O remains zero up to radially outward few nearest neighbour locations in both cases as illustrated in Figure 3(b) and 3(c). In both cases, BNC for cationic species, either Ga or Al in $Ga_2O_3$ or $Al_2O_3$, respectively, is higher in the radially outward positions (d>0) surrounding the dopants located at the center (d = 0). The high Ga or Al concentrations surrounding the Si atoms indicates, dopant Si atoms are more likely occupying in Ga site in $Ga_2O_3$ as in Figure 3(b) or Al site in $Al_2O_3$ as in Figure 3(c). At the reference dopant atoms' position (d = 0), BNC for Ga and Al drops to zero as shown in

Fig. 3(b) and 3(c), respectively. This implies the presence of cation vacancies ($V_{Ga}$ and $V_{Al}$ in $Ga_2O_3$ and $Al_2O_3$, respectively) and dopants are interacting with these cationic vacancies [13,16]. Such specific cationic site occupancy of Si and its interaction with cation vacancies would lead to the formation of complex defects such as $V_{Ga}$-$Si_{Ga}$ or $V_{Al}$-$Si_{Al}$ as observed in the case of GaN or GaAs and AlN [13,15-17]. Considering the RDF results showing cationic substitution for x = 0 ($Ga_2O_3$) and x = 1.0 ($Al_2O_3$) as a reference, we investigated cation (Ga or Al) site substitution by dopant in rest of the alloy composition with 0 < x < 100 in the next part of this paper.

Since the alloy composition is varying with Al content from x = 0-100%, we extended our analysis to other $(Al_xGa_{1-x})_2O_3$ layers with x > 0 and x < 1.0. In these layers, there are both Ga and Al sites that Si dopants can occupy. The RDF analysis for $(Al_xGa_{1-x})_2O_3$ films with x = 10%, 20%, 30%, 40%, 50%, 60% and 80% are illustrated in Figure 4(a-i). At low Al content, x = 10-20%, RDF analysis shows that in the proximity of the center Si atoms up to few nearest neighbour locations, BNC for Ga is higher while that for Al is low (BNC~0 for Al) which indicates Si is likely to occupy on Ga site as depicted in Figure 4 (a-c). At the center Si atoms (distance, d = 0), BNC for Ga drops to zero, implying the presence of $V_{Ga}$ (Figure 4(a-c)). This suggests that the dopant Si atoms are interacting with these $V_{Ga}$ that would lead to the formation of vacancy-interstitial complex defects such as $V_{Ga}$-$Si_{Ga}$ [13,17]. Within this Al composition range (x = 10-20%), $(Al_xGa_{1-x})_2O_3$ films are single monoclinic β-phase stable [3,28], so our analysis recommends, in β-$(Al_xGa_{1-x})_2O_3$, Si doping is preferably occupying on Ga sites which may lead to the formation

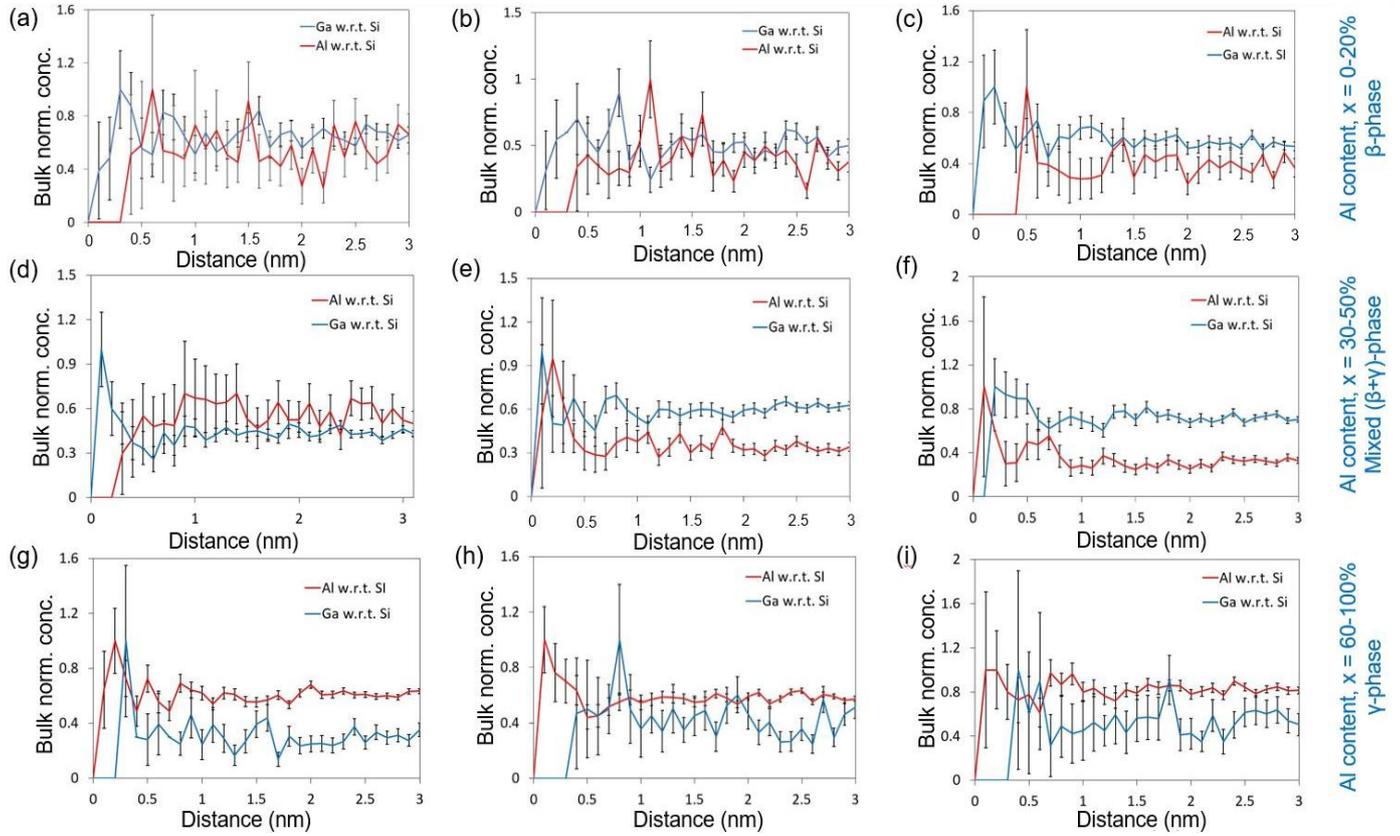

Figure 4. Radial distribution function results in each $(Al_xGa_{1-x})_2O_3$ layers showing Si is occupying (a-c) Ga site at $(Al_xGa_{1-x})_2O_3$ at x = 0.10-0.20; (d-f) Ga or Al site at $(Al_xGa_{1-x})_2O_3$ at x = 0.30-0.50; (g-i) Al site at $(Al_xGa_{1-x})_2O_3$ at x = 0.60-0.80.

of complex $V_{Ga}$-$Si_{Ga}$ defects due to the interaction with $V_{Ga}$ vacancies.

The RDF for $(Al_xGa_{1-x})_2O_3$ with x = 30-50% are shown in Fig. 4(d-f). In Fig. 4(d), both Al and Ga BNC are higher from d = 0 and which site Si is occupying cannot be determined conclusively. In some cases, it is observed that from the origin (*i,e,* reference Si atoms at d = 0) towards a radially outward direction, BNC of Al is zero up to few nearest neighbour locations while BNC for Ga in locations surrounding d = 0 is higher (Fig. 4(e)). This result is similar to what was observed in the case of Al content, x = 10-20% and suggests that Si is occupying on Ga sites. At d=0, BNC for Ga drops to zero and suggest that $V_{Ga}$ is present. However, there is another case when high Al concentration is observed in the immediate surrounding positions of the centering Si atoms followed by BNC~0 for Ga up to few nearest neighbours indicating Al site occupancy of Si dopants. BNC for Al at d=0 is implying possible presence of Al vacancy ($V_{Al}$) with which dopant Si would interact and form defect complexes [13,17]. These contradictory results within Al content of x = 30-50% layers indicate dopant Si atoms can occupy either Ga or Al site which can contribute to either $V_{Ga}$-$Si_{Ga}$ or $V_{Al}$-$Si_{Al}$ defect complex formation. We believe, crystallinity degradation due to the presence of mixed (β+γ)-phases in $(Al_xGa_{1-x})_2O_3$ at x = 30-50% [22] is playing a pivotal role in such conflicting RDF results in these layers. Such phase segregation results in undetermined Al or Ga site occupancy of dopant atoms depending on these dopants are in the Al-rich or Al depleted regions.

It is reported in our previous work, that the $(Al_xGa_{1-x})_2O_3$ layers with higher Al content, x = 60-80% regains its single phase (γ-phase) crystalline structure [22]. The RDF analysis in these layers is illustrated in Fig. 4(g-i). From the RDF data, it is observed that surrounding the reference Si atoms up to few nearest neighbours, Al concentration is higher compared to Ga concentrations (BNC for Ga ~0). This refers to Si is substituting on Al site unlike Ga at low Al content $(Al_xGa_{1-x})_2O_3$ films. At the origin (d = 0; at reference Si) BNC for Al drops to zero. Our hypothesis is, at high Al content, the formation energy of $V_{Al}$ is lower [29] thereby this kind of defects becomes dominant over $V_{Ga}$ with which dopant Si atoms would interact. Similar phenomenon was observed in the case of AlGaN with high Al content (x > 60%) [16]. Within this Al range, dopant Si atoms more likely to occupy on Al sites and interact with $V_{Al}$ defects that leads to the formation of $V_{Al}$-$Si_{Al}$ complex defect structures [13,17]. This specific cationic site occupancy by dopant Si contributes to the formation of cation specific defect complexes ($V_{III}$-$Si_{III}$) [30]. At low Al content (x<0.30), such defect complexes are mostly $V_{Ga}$-$Si_{Ga}$ which would drive the compensating knee (lower mobility because of reduced free electrons at high doping level) in $(Al_xGa_{1-x})_2O_3$ films. On the contrary, $V_{Al}$-$Si_{Al}$ complex defect would be the dominant one to tailor the compensating effect in this film at high Al content (x>0.50) resulting in degraded device efficiencies [19]. This information will be valuable for doping design in $(Al_xGa_{1-x})_2O_3$ towards high performance devices.

## 4. Conclusion

We investigated the dopant interaction in Si doped $(Al_xGa_{1-x})_2O_3$ films with varying Al content over a wide range of Al composition using APT. Moderate dopant density in the range of $10^{18}$ cm$^{-3}$ with an almost uniform doping profile in each $(Al_xGa_{1-x})_2O_3$ layers irrespective of the Al content was achieved. In single phase β-$(Al_xGa_{1-x})_2O_3$ crystalline layers (Al< 0.3), Si is found to occupy Ga sites which could be contributing to the formation of $V_{Ga}$-$Si_{Ga}$ defect complexes. When $(Al_xGa_{1-x})_2O_3$ layers are of mixed phase (0.3<Al<0.6), Si prefers to occupy either Ga or Al cationic site. The reason could be attributed to the presence of different chemical phases in these layers that results in Al-rich or Al-depleted regions. On the contrary, at high Al content (Al>0.6) single phase γ-$(Al_xGa_{1-x})_2O_3$ films, Si prefers to occupy Al site instead of Ga. This Al site occupancy can account for the formation of $V_{Al}$-$Si_{Al}$ complex arising from an increasing number of $V_{Al}$ sites. This specific cationic site occupancy by dopant Si would account for compensation knee (reduced mobility despite high doping level) resulting from the formation of cation specific III-defect complex ($V_{III}$-$Si_{III}$). This understanding of dopant behavior in Si doped $(Al_xGa_{1-x})_2O_3$ films would be highly significant in terms of doping wide bandgap semiconductors for high power electronics as well as solar blind photodetectors.

## Acknowledgements

Bhuiyan, Feng, Zhao acknowledge the funding support from the Air Force Office of Scientific Research FA9550-18-1-0479 (AFOSR, Dr. Ali Sayir). Feng and Zhao also acknowledge partial support from the National Science Foundation (1810041). The authors also acknowledge the contribution of Menglin Zhu (The Ohio State University) in atom probe specimen preparation.

## References

[1] Ueda N, Hosono H, Waseda R, and Kawazoe H, 1997, Appl. Phys. Lett. **71(7)**, 933.

[2] Gillet E and Ealet B, Surf. Sci., 1992, **273 (3)**, 427.

[3] Krueger B W, Dandeneau C S, Nelson E M, Dunham S T, Ohuchi F S, and Olmstead M A, 2016, J. Am. Ceram. Soc., **99 (7)**, 2467.

[4] Zhang Y, Neal A, Xia Z, Joishi C, Johnson J M, Zheng Y, Bajaj S, Brenner M, Dorsey D, Chabak K, Jessen G, Hwang J, Mou S, Heremans J P, and Rajan S, 2018, Appl. Phys. Lett. **112**, 173502.

[5] Vaidya A, Sarker J, Zhang Y, Lubecki L, Wallace J, Poplawsky J D, Sasaki K, Kuramata A, Goyal A,

Gardella J A, Mazumder B, and Singisetti U, J. Appl. Phys., 2019, **126**, 095702.

[6] Yuan S H, Wang C C, Huang S Y, Wuu D S, IEEE Electron Device Lett., 2018, **39(2)**, 220.

[7] Dong L, Jia R, Xin B, Peng B, and Zhang Y, Scientific Reports, 2017, **7**, 40160.

[8] Hassa A, Wenckstern H v, Vines L, and Grundmann M, ECS J. Solid State Sci. Technol., 2019, **8(7)**, Q3217.

[9] Farzana E, Ahmadi E, Speck J S, Arehart A R, and Ringel S A, J. Appl. Phys., 2018, **123**, 161410.

[10] Feng Z, Bhuiyan A F M A U, Karim M R, Zhao H, Appl. Phys. Lett., 2019, **114**, 250601.

[11] Bhuiyan A F M A U, Feng Z, Johnson J M, Chen Z, Huang H -L, Hwang J, and Zhao H, Appl. Phys. Lett., 2019, **115**, 120602.

[12] Ranga P, Rishinaramangalam A, Varley J, Bhattacharyya A, Feezell D, and Krishnamoorthy S, 2019, Appl. Phys. Express, **12**, 111004.

[13] Chichibu S F, Miyake H, Ishikawa Y, Tashiro M, Ohtomo T, Furusawa K, Hazu K, Hiramatsu K, and Uedono A, J. Appl. Phys., 2013, **113**, 213506.

[14] Gogova D, Wagner G, Baldini M, Schmidbauer M, Irmscher K, Schewski R, Galazka Z, Albrecht M, Fornari R, J. Cryst. Growth, 2014, **401**, 665.

[15] Chichibu S, Iwai A, Nakahara Y, Matsumoto S, Higuchi H, Wei L, and Tanigawa S, J. Appl. Phys., 1993, **73**, 3880.

[16] Götz W, Johnson N M, Chen C, Liu H, Kuo C, and Imler W, 1996, Appl. Phys. Lett. **68**, 3144.

[17] Uedono A, Tenjinbayashi K, Tsutsui T, Shimahara Y, Miyake H, 2012, J. Appl. Phys. **111**, 013512.

[18] Zhang Z, Farzana E, Arehart A R, and Ringel S A, Appl. Phys. Lett., 2016, **108**, 052105.

[19] Harris J S, Baker J N, Gaddy B E, Bryan I, Bryan Z, Mirrielees K J, Reddy P, Collazo R, Sitar Z, and Irving D L, Appl. Phys. Lett., 2018, **112**, 152101.

[20] Kelly T F, Larson D J, Thompson K, Alvis R L, Bunton J H, Olson J D, and Gorman B P, Annu. Rev. Mater. Res., 2007, **37**, 681.

[21] Perea D, Arslan I, Liu J, Nat. Commun., 2015, **6,** 7589.

[22] Bhuiyan A F M A U, Feng Z, Johnson J, Huang H -L, Sarker J, Zhu M, Karim M R, Mazumder B, Hwang J, and Zhao H, 2020, APL Mater., **8**, 031104.

[23] Thompson K, Lawrence D, Larson D J, Olson J D, Kelly T F, and Gorman B, 2007, *Ultramicroscopy*, **107(2-3)**, 131.

[24] A. Devaraj, M. Gu, R. Colby, P. Yan, C.M. Wang, J.M. Zheng, J. Xiao, A. Genc, J.G. Zhang, I. Belharouak, D. Wang5, K. Amine5 & S. Thevuthasan, Nature Communications, 6, 8014, (2015).

[25] Devaraj A, Perea D E, Liu J, Gordon L M, Prosa T J, Parikh P, Diercks D R, Meher S, Kolli R P, Meng Y S and Thevuthasan S, Int. Mater. Rev., 2018, **63(2)**, 68.

[26] Zhou J, Odqvist J, Thuvander M, and Hedström P, Microsc. Microanal., 2013, **19**, 665.

[27] Schmidt J, Peng L, Poplawsky J D and Weckhuysen B M, Angew. Chem. Int. Ed., 2018, **57 (33)**,10422.

[28] Kaun S W, Wu F, and Speck J S, J. Vac. Sci. Technol. A, 2015, **33**, 041508.

[29] Stampfl C and Walle C G V D, Phys. Rev. B, 2002, **65**, 155212.

[30] Rass J, (Ed.), 2015, III-Nitride Ultraviolet Emitters-Technology and Applications, Springer International Publish.